\documentclass[pre,aps,twocolumn]{revtex4}
\usepackage{epsfig}
\usepackage{bm} 
\newcommand{\B}[1]{{\bm{#1}}}

\newcommand{\Onecol} {\begin{widetext} \onecolumngrid} 
\newcommand{\Twocol} {\end{widetext} \twocolumngrid} 

\newcommand{\be}{\begin{equation}}
\newcommand{\ba}{\begin{array}}
\newcommand{\bea}{\begin{eqnarray}}
\newcommand{\bfi}{\begin{figure}}
\newcommand{\ee}{\end{equation}}
\newcommand{\ea}{\end{array}}
\newcommand{\eea}{\end{eqnarray}}
\newcommand{\efi}{\end{figure}}
\newcommand{\dv}{\delta_r v}

\newcommand{\Bv}{\B{v}}
\newcommand{\Bx}{\B{x}}
\newcommand{\Bf}{\B{f}}

\begin{document} 
\bibliographystyle{prsty}
\title{Dynamical scaling and intermittency in shell models of turbulence}
\author{Roberto Benzi$^{1}$, Luca Biferale$^{1}$ and Mauro Sbragaglia$^{1}$} 
\affiliation{$^1$ Dipartimento di Fisica and INFN, Universit\`a ``Tor
Vergata", Via della Ricerca Scientifica 1, I-00133 Roma, Italy.} 
\begin{abstract} 
We introduce a  model for  the turbulent energy cascade
aimed at studying the effect of dynamical scaling on intermittency. 
In particular, we show
that  by slowing down the 
energy transfer mechanism for fixed energy flux,
intermittency decreases and eventually disappears. This result supports the 
conjecture that
intermittency 
can be observed only if energy is flowing towards faster and faster 
scales of motion.

\vskip 0.2cm 
\end{abstract} 
\maketitle 
Turbulent flows show very different behaviour at changing the
embedding physical dimension.  The most spectacular change is in the
reversal of the energy flux, from large to small scales in $3d$ and
viceversa in $2d$ \cite{fri95}. Moreover, the three-dimensional
forward energy flux is strongly intermittent while the two-dimensional
inverse energy transfer is almost Gaussian \cite{bof00}. The presence
of strong anomalous fluctuations in the $3d$ flux is believed to be
connected to the existence of a hierarchical organization in the eddy
turn over times, $\tau_r$, at different scales. Simple
dimensional arguments predict the eddy turn over time of velocity
fluctuations at scale $r$ in the inertial range to be of the order of
$\tau_r \sim r^{2/3}$. For three dimensional turbulence, the 
scenario is the one of an energy cascade
from slow (large eddies) to fast (small eddies), without the
possibility for fluctuations at different scales to {\em equilibrate}. 
This mechanism should be at the origin of the burst-like
structure in the forward energy transfer mechanism: any unusual
fluctuation in the energy content at a given  scale propagates 
to smaller and smaller scales until it is re-adsorbed by viscous  mechanism. 
Hence, this is the cause of
strong inhomogeneity in the spatio-temporal statistical properties of
the energy transfer process.  The same qualitative arguments capture
the absence of intermittency in the inverse $2d$ process, where energy
flows from fast to slow modes, allowing the receiving modes to feel
only the {\em mean} fluctuations of the unstable ``mother''
eddy. Intermittency 
in the related problem  of scalar passive/active quantities advected
by turbulent flows has been connected to the
 existence of a ``dissipative anomaly'', i.e. the presence of a
 non-vanishing  scalar dissipation in the limit of small molecular
 diffusivity. Intermittency in the latter case has a natural
 Lagrangian interpretation connected to  the
 existence of   particles which separate even if starting in
 coinciding points \cite{rev,sca_pas}.   Such a mechanism is absent in inverse cascade,
 pointing toward the conclusion that inverse cascade regimes cannot be
 intermittent. Here we intend to investigate a similar issue on a purely 
 Eulerian base. 

Intermittent fluctuations are usually related to the scaling
properties of the Navier Stokes equations \cite{fri95}. Let us denote
by $\Bv( \Bx,t)$ the velocity field of a turbulent flow satisfying the
Navier Stokes equations:
\begin{equation}
\partial_t \Bv + \Bv \bullet \nabla \Bv = - \nabla \pi + \nu \Delta \Bv + \Bf
\label{NS}
\end{equation}
where $\pi = p/\rho$ ($p$ being the pressure and $\rho$ the density),
$\nu$ is the kinematic viscosity of the flow and $\Bf$ is the (large
scale) external forcing. Equations (\ref{NS}) show the remarkable
property to be invariant under the scale transformation
\begin{equation}
\Bx \rightarrow \lambda \Bx \ , \ \ \Bv \rightarrow \lambda^h \Bv
\ , \ \ t \rightarrow \lambda^{1-h} t \ , \ \ \nu \rightarrow
\lambda^{1+h} \nu.
\label{h}
\end{equation}
Note that (\ref{h}) implies for the energy dissipation: $\epsilon
\rightarrow \lambda^{3h-1} \epsilon$, i.e. $\epsilon$ is constant if
$h=1/3$ as predicted by the K41 theory. In the multifractal theory of
turbulent flows, $h$ is supposed to be a fluctuating quantity,
although on the average the rate of energy dissipation is constant
\cite{fri95}. The above scenario is complemented by the assumption
that the statistical fluctuations are described by a scaling invariant
probability distribution $P_h \sim \lambda^{3-D(h)}$ where $D(h)$ can
be interpreted as the fractal dimension related to the fluctuations
$\lambda^h$. Regardless of the geometrical interpretation of $D(h)$,
the basic physical question is which is the mechanism determining the
fluctuations of $h$ and why there are fluctuations in the energy
flux. The above, scale invariant, scenario of the non
equilibrium statistical properties of turbulent flows suggests a
simple although not trivial picture of intermittency, related to the
energy cascade mechanism.  At scale $r$, the amount of kinetic energy
due to turbulent fluctuations can be estimated as $(\delta_r v)^2$,
where $\delta_r v = v(x+r)-v(x)$, and where we neglect vectorial
indexes for simplicity. We expect that the rate of energy flux at
scale $r$, \ denoted by $\epsilon_r $, is of the order of $(\delta_r
v)^2 / \tau_r$. Because energy transfer is due to non linear
interactions, $\tau_r$ can be estimated as $r/\dv$. Thus we obtain
$(\delta_r v)^3 \sim r \epsilon(r)$ which, as expected, is invariant
under (\ref{h}).  As a result, the energy transfer statistics is
strongly correlated to the fluctuations on the energy contents at
different scales, $(\delta_r v)^2$, and to their dynamical properties,
$\tau_r$. Moreover, the presence of strong intermittent fluctuations in the 
energy contents at different scales, reflects, via the equation of
motions,  into non-trivial fluctuations of the local eddy turn
over times. Indeed, previous theoretical and numerical works have
demonstrated that spatial and temporal properties of the energy cascade mechanism are
strongly correlated \cite{proc,bif_pd}.  

The previous phenomenological  arguments suggest  that intermittency
can be observed  only if energy is flowing towards faster
scales of motion. It is therefore tempting to argue that by decreasing the scaling exponents of 
the eddy turnover times along the cascade, i.e. by slowing down the 
energy transfer mechanism for fixed energy flux,
intermittency should decrease and eventually disappear.
Our aim in this letter is to provide clear evidence that the above
conjecture  holds. \\

Our analysis will be performed in the 
framework of 
 shell models of turbulence
(see \cite{fri95,bohr,bif03} and references therein). The motivation to use shell models as a
possible surrogate of the Navier-Stokes dynamics is twofold. First,
shell models proved to be very successful in reproducing many of the
statistical features of both $2$ and $3$ dimensional turbulent flows,
being at the same time much easier to simulate numerically. Second,
they are flexible enough to allow a structural change in their
equations of motion which will allow us to directly probe the
importance of time dynamics in fixing the intermittent properties of
the energy transfer process (see below).
 
 In a shell model, the basic variable describing the 'velocity
field' at scale $r_n = \Lambda^{-n} r_0 \equiv k_n^{-1}$, is a complex
number $u_n$ satisfying a suitable set of non linear equations. There
are many version of shell models which have been introduced in
literature (see \cite{bif03} for a recent review). Here we choose the
one proposed in \cite{lvo98} which is an improved version of the
so-called GOY model \cite{yam88b,gio}
\begin{eqnarray}
&& \frac{d u_n}{dt} =   i k_n [\Lambda u_{n+1}^*u_{n+2} 
  + b u_{n-1}^*u_{n+1} -c \Lambda^{-1} u_{n-2}u_{n-1} ] \nonumber \\
&& - \nu k_n^2 u_n + f_n  
  \label{sabra}
\end{eqnarray}
where $\Lambda=2$, $c= -(1+b)$ and $f_n$ is an external forcing.  In
shell models, we can associate $\delta_r v$ to $u_n$.  Clearly
equation (\ref{sabra}) satisfies the scaling (\ref{h}).  The
important point on shell models like (\ref{sabra}) is that the
statistical properties of intermittent fluctuations, computed either
using $u_n$ or the instantaneous rate of energy dissipation, are in
close {\em qualitative} and {\em quantitative} agreement with those
measured in laboratory experiments, for homogeneous and isotropic
turbulence.  Thus, shell models provide an useful tool to investigate
in a simple way the physical consequences of scaling (\ref{h}) and
intermittency. Moreover, at variance from $3d$ Navier-Stokes
equations, the computational complexity grows with Reynolds only as
$Re^{1/2}$, allowing for reliable numerical studies also at very high
Reynolds numbers.

It is easy to realize that also in shell models, $\tau({k_n})$ goes as
  $k_n^{h-1}$ as predicted by (\ref{h}), i.e. the characteristic time
  for energy transfer decreases quite fast as the scale $k_n^{-1}$ is
  decreasing. If $\tau(k_n)$ is independent of $k_n$, the
  constant-flux argument
  would suggest energy equipartition among all scales, i.e. $(\delta_r
  v)^2 \sim \epsilon_r$. To be more specific, by constraining ourselves on
  shell models, let us modify (\ref{sabra}) in such a way that
  there still exists an average Reynolds independent rate of energy
  dissipation $\epsilon = \langle \nu \sum_n k_n^2 |u_n|^2 \rangle$
  {\em while} the statistical properties are invariant under the
  scaling transformation:
\begin{equation}
k_n \rightarrow \lambda^{-1} k_n\ , \ \ u_n \rightarrow \lambda^{h} u_n \ , \
\ t \rightarrow \lambda^{x-h} t \ , \ \ \nu \rightarrow \lambda^{1+h}
\nu
\label{hh}
\end{equation}
where $x$ is a fixed parameter in the equations. Note that (\ref{hh})
to hold, we should require that the non linear terms scale as
$\lambda^{2h-x}$. Note also that $\epsilon \sim \nu (\nabla v)^2 \sim
\lambda^{3h-1}$, thus we do not change the constrain on the energy
flux.  

In order to satisfy the scaling (\ref{hh}), 
we introduce the following equations:
\begin{eqnarray}
&&\frac{d u_n}{dt} = i k_0^{1-x}k_n^x [u_{n+1}^*u_{n+2} + b
  u_{n-1}^*u_{n+1} \nonumber \\ && +\frac{1+b}{2} u_{n-2}u_{n-1} ] -
  \nu k_n^{1+x} k_0^{1-x}u_n + f_n
  \label{sabrax}
\end{eqnarray}
where $k_0$ is the largest scale in the system. Clearly,  we obtain 
the old shell model for $x=1$, while
a direct inspection of (\ref{sabrax}) shows that (\ref{hh}) is satisfied for any $x$.  
Moreover, for $\nu=0$, the generalized energy,  
$Q = k_0^{x-1}\Sigma_n k_n^{1-x}|u_n|^2$, is conserved by non liner
interactions. Thus,  
in a statistically stationary regime we have the 'energy' budget:
$$ \frac{1}{2}\frac{dQ}{dt} = 0 = P - \nu \sum_n k_n^2 |u_n|^2
$$ where $P\propto Real(\langle \sum_n k_n^{1-x} u_n^*f_n \rangle)$ is
the energy input.  Note that in this shell model energy fluctuations
at scale $k_n$ are of the order $Q_n = k_0^{x-1}k_n^{1-x}|u_n|^2$
while the eddy-turn-over time for the energy transfer is of the order
\begin{equation}
\tau(k_n) \equiv 1/(k_0^{x-1}k_n^xu_n).
\label{taux}
\end{equation}
Thus the energy flux through wavenumber $k_n$ can be proved rigorously
to be of the order of $\Pi_n = k_n |u_n|^3$ as for the original shell
model. Therefore, for all $x$,  by keeping $\langle \Pi_n \rangle =
const.$ we can achieve a constant energy flux:
\begin{equation}
\langle |u_n|^3 \rangle  \sim k_n^{-1}.
\label{45x}
\end{equation}
In summary,
 while for all $x$ the shell model (\ref{sabrax}) produces the same average energy flux, its dynamics is
characterized by a different time-scaling properties which changes
the  eddy-turn-over  hierarchy  in the inertial range. 
In particular for small $x$, 
say $x \sim 1/3$, by combining (\ref{taux}) and (\ref{45x}) we
should expect all eddy turn over times to be of the same order in the
inertial range and therefore  the energy to reach a quasi-equipartion
state for fixed value of
$\epsilon$.
The shell model provides us the interesting possibility to study
intermittency as function of $x$, i.e. as a function of the {\em
dynamical} scaling (\ref{hh}) and (\ref{taux}). Intuitively, one can
imagine that decreasing $x$ from its Navier-Stokes value $x=1$
induces a smoother and smoother energy transfer process towards
smaller scales, fluctuations among different scales tend to
equilibrate each other and, consequently, non-Gaussian fluctuations
are depleted. Thus, as a function of $x$ the shell model
(\ref{sabrax}) should
exhibit a kind of phase transition from 'strong' intermittent
fluctuations at $x=1$ to Gaussian non intermittent fluctuations at $x$
small.  In this paper we support the above conjecture by numerical
simulations of the  model (\ref{sabrax}). Moreover, we will give a theoretical
argument to estimate the value of $x_c$ below which intermittency
starts to be depleted.

\begin{figure}[hbt]
\includegraphics[scale=0.68]{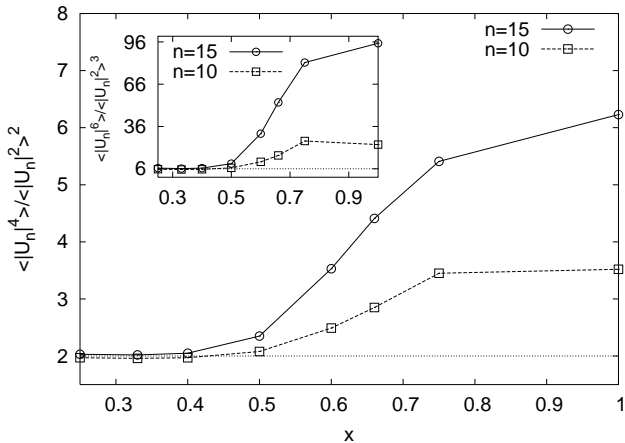}
\caption{The value of $G_{2p}(k_n) = \langle |u_n|^{2p}
\rangle/\langle|u_n|^2\rangle^p$ for $p=2$ and $p=3$ (in the inset)
as function of $x$ and for two different values of $k_n$, namely
$n=10$ and $n=15$. The predicted value for the Gaussian Statistics (horizontal dotted line) is reported for comparison.}
\label{fig1}
\end{figure}

In order to keep $\epsilon$ constant, regardless of the value of $\nu$
and $x$, we use the forcing $f_n = A_n/u_n^*$ for $n=1,2$ and $f_n=0$ for $n>2$.
We have performed a series of numerical simulations for different
values of $x$, keeping the parameter $b=-0.4$ fixed.  As a
measure of intermittency, we compute the generalized kurtosis
$G_{2p}(k_n) = \langle |u_n|^{2p}
\rangle/\langle|u_n|^2\rangle^p$. Figure (\ref{fig1}) shows the value of
$G_4(k_n)$ for different values of $k_n$ as a function of $x$, while in the 
inset we show the value of $G_6(k_n)$. 
As one clearly see, for $x$
smaller than $0.5$ a sharp decrease of $G_4$ and $G_6$ is observed. For small values of $x$
both $G_4$ and $G_6$
become equal to their Gaussian value.
The above results tell us that intermittency is depleted for small $x$
in agreement with our intuitive argument. Also, within the numerical
error bars, for $x> x_c \sim 0.5$, intermittent fluctuations seem to
weakly depend on $x$.  

We want now to understand why the transition to
no-intermittency fluctuations is observed for relatively large value
of $x$, namely for $x \sim 0.4$. As we shall see in the following,
$x_c = 1/3 + \Delta x$ where $\Delta x$ is due to intermittency
correction.  The argument goes as follows. Let us consider two scales
$k_n$ and $k_{n+m}$. The corresponding times for the energy transfer are
$\tau(k_n) \sim k_n^{h}$ and $\tau(k_{n+m}) \sim k_{n+m}^{h^{'}}$,
where $h$ and $h^{'}$ are the values of the 'local' fluctuations
of $u_n \sim k_n^{-h}$ and $u_{n+m} \sim k_{n+m}^{-h^{'}}$
respectively. For $x=1$ the probability that the ratio $\chi_m =
\tau(k_{n+m})/\tau(k_n)$ is larger than $1$ is extremely small, already for
$m=1,2$. When $x$ becomes smaller than $1$ the probability
$P(\chi_m>1)$ start growing, even for small $m$. If $\chi_m$ is much
larger than $1$, then the energy transfer from scale $k_n$ to scale
$k_{n+m}$ is stopped and energy tends to equipartition. In order to
estimate $x_c$, let us compare, for a given scale separation $k_m$, the ratio between 
two eddy turn over times $T_m(q) = \langle(\chi_m)^q\rangle^{1/m}$:
\begin{eqnarray}
&& T_m(q) \sim \left(\int dh k_m^{(qh-qx-(3-D(h))}\right)^{1/m} \sim
k_m^{\frac{(-qx- \zeta(-q))}{m}} \nonumber \\
&& \zeta(q) = inf_h [qh+3-D(h)].
\end{eqnarray}  

\begin{figure}[hbt]
\includegraphics[scale=0.68]{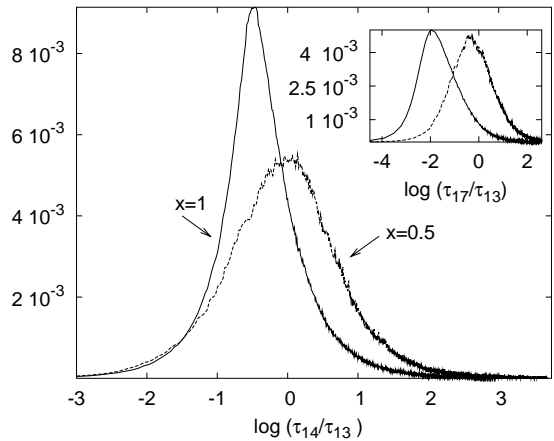}
\caption{Probability distribution of $log(\tau(k_{14}) / \tau(k_{13}))$ for
two different values of $x$, namely $x=1$ and $x=0.5$. The theory
presented in the paper predicts that intermittency is depleted at
$x=0.5$ because $\tau(k_{n+1}) \sim \tau(k_{n})$. In the inset we show the
probability distribution of $log( \tau(k_{17})/\tau(k_{13}))$}
\label{fig2}
\end{figure}
The above expression tells us that $T_m(q)>1$ for $x<x_c(q)=
 -\zeta(-q)/q$. Let us note that, because of convexity properties of
 $\zeta(q)$, $x_c(q)$ is an increasing function of $q$.  Thus, there is
 not a single value of $x$ below which intermittency is depleted,
 rather the transition to equipartition is continuous, different
 moments of the eddy turn over ratios behave in slightly different
 way. As a simple guide line  the transition
 region is $[x_c(1),x_c(2)] = [0.4,0.42]$, estimated
  by using the $D(h)$ curve which fits  the
 $\zeta(q)$ exponents as given by the She-Leveque formula \cite{she}.
\begin{figure}[hbt]
\includegraphics[scale=0.68]{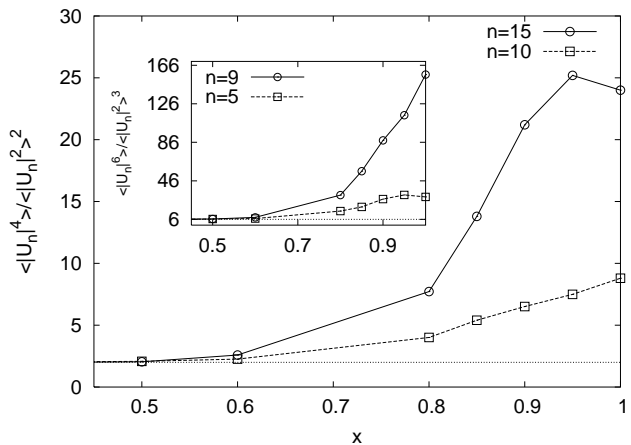}
\caption{Same quantities as in figure (\ref{fig1}) for the 
model with parameter $b=-0.8$. The increase of intermittency at $x=1$
increases the critical value of $x$ for which {\em thermodynamic
equilibrium} is attained.}
\label{fig3}
\end{figure}
A direct numerical investigation gives further support to our previous
argument. In figure (\ref{fig2}) we show the probability density
function of the ratio $\tau(k_{14})/\tau(k_{13})$, $\tau(k_n)$ being the
instantaneous eddy turnover time at shell $k_n$. For $x=1$, we expect
that $\tau(k_{14}) < \tau(k_{13})$ with probability close to $1$, while,
according to our estimate, we expect that for $x=0.5$
 $\tau(k_{14}) \sim \tau(k_{13})$ and intermittency is depleted.  This is
clearly shown in figure (\ref{fig2}), where in the insert we also show
the probability density function of $log(\tau(k_{17})/ \tau(k_{13}))$. Hence, we
can conclude that for all scales $k_n$ the eddy turnover times become
almost equal producing a kind of {\em thermodynamic} equilibrium in
the generalized energy.

As a further check of our argument, we consider the shell model
 (\ref{sabrax}) for
 $b=-0.8$. In this case, the model at $x=1$ shows larger intermittent
 correction with respect to the case $b=-0.4$ previously
 considered. According to our argument for the transition to occur,
 we should expect that for $b=-0.8$, depletion of intermittency takes
 place for larger values of $x$, which is indeed the case as shown in
 figure (\ref{fig3}).

In summary, we have presented the following main results: (i) we
introduced a new version of the shell model which satisfies a
generalization of the dynamical scaling (\ref{hh}); (ii) we have
proposed a simple, although non trivial, argument for understanding
how intermittency can depend on the scaling properties of the eddy
turnover time; (iii) we have shown, by numerical simulations, that our
argument is correct; (iv) we have provided a multifractal estimate of
the critical value, $x_c$ where  intermittency should disappear.\\
We thank F. Toschi for discussion in a early stage of this work. 


\begin{thebibliography}{10}
\bibitem{fri95}
U. Frisch, {\em Turbulence: The legacy of A.N. Kolmogorov} (Cambridge
  University Press, Cambridge, 1995).
\bibitem{bof00} G. Boffetta, A. Celani, and M. Vergassola,
Phys. Rev. E {\bf 61}, R29 (2000).
\bibitem{rev} G. Falkovich, K. Gawedzki and M. Vergassola 
       Rev. Mod.  Phys. {\bf 73} 913 (2001).       
\bibitem{sca_pas} A. Celani, M. Cencini, A. Mazzino  and M. Vergassola
  New J. Phys. {\bf  6}, 72 (2004).
\bibitem{proc} V.S. L'vov, E. Podivilov and I. Procaccia. Phys. Rev. E {\bf
  55} 7030  (1997).
\bibitem{bif_pd}  L. Biferale, G. Boffetta, A. Celani and F. Toschi. 
Physica D {\bf 127} 187 (1999).
\bibitem{bohr}
T. Bohr, M.~H. Jensen, G. Paladin, and A. Vulpiani, {\em Dynamical Systems
  Approach to Turbulence} (Cambridge University Press, Cambridge, 1998).
\bibitem{bif03}
L. Biferale, Ann. Rev. Fluid. Mech. {\bf 35},  441  (2003).
\bibitem{lvo98}
V.S. L'vov, E. Podivilov, A. Pomyalov, I. Procaccia, and D. Vandembroucq, Phys.
  Rev. E {\bf 58},  1811  (1998).
\bibitem{yam88b}
M. Yamada and K. Ohkitani, Phys. Rev. Lett. {\bf 60},  983  (1988).
\bibitem{gio} M.H. Jensen, G. Paladin and A. Vulpiani, 
  Phys. Rev. A. {\bf 43} 798 (1991).
\bibitem{she} Z.S. She and E. Leveque, Phys. Rev. Lett. {\bf 72} 336 (1994). 
\end{thebibliography}

\end{document}